\documentclass[12pt]{article}
\usepackage[a4paper,total={18cm,27cm}]{geometry}
\usepackage{graphicx}
\usepackage{authblk}
\usepackage{amsmath}
\usepackage[utf8]{inputenc}
\usepackage{hyperref}
\usepackage{verbatim}
\usepackage{color,soul}
\usepackage{caption}
\usepackage{xcolor}
\usepackage{framed}
\colorlet{shadecolor}{yellow!20}

\begin{document}


\title{ 
Extracting continuous sleep depth from EEG data\\ without machine learning 
}




\author[1]{Claus Metzner}
\author[1,2]{Achim Schilling}
\author[3]{Maximilian Traxdorf}
\author[1]{Holger Schulze}
\author[1]{Konstantin Tziridis}
\author[1,2,4]{Patrick Krauss}
\affil[1]{\small Neuroscience Lab, Experimental Otolaryngology, University Hospital Erlangen, Germany}
\affil[2]{\small Cognitive Computational Neuroscience Group, Friedrich-Alexander-Universität Erlangen-Nürnberg, Germany}
\affil[3]{\small 
Department of Otorhinolaryngology, Head and Neck Surgery, Paracelsus Medical University, Nürnberg, Germany}
\affil[4]{\small Pattern Recognition Lab, Friedrich-Alexander-Universität Erlangen-Nürnberg, Germany}

\maketitle


\begin{abstract}
The human sleep-cycle has been divided into discrete sleep stages that can be recognized in electroencephalographic (EEG) and other bio-signals by trained specialists or machine learning systems. It is however unclear whether these human-defined stages can be re-discovered with unsupervised methods of data analysis, using only a minimal amount of generic pre-processing. Based on EEG data, recorded overnight from sleeping human subjects, we investigate the degree of clustering of the sleep stages using the General Discrimination Value as a quantitative measure of class separability. Virtually no clustering is found in the raw data, even after transforming the EEG signals of each thirty-second epoch from the time domain into the more informative frequency domain. However, a Principal Component Analysis (PCA) of these epoch-wise frequency spectra reveals that the sleep stages separate significantly better in the low-dimensional sub-space of certain PCA components. In particular the component $C_1(t)$ can serve as a robust, continuous 'master variable' that encodes the depth of sleep and therefore correlates strongly with the 'hypnogram', a common plot of the discrete sleep stages over time. Moreover, $C_1(t)$ shows persistent trends during extended time periods where the sleep stage is constant, suggesting that sleep may be better understood as a continuum. These intriguing properties of $C_1(t)$ are not only relevant for understanding brain dynamics during sleep, but might also be exploited in low-cost single-channel sleep tracking devices for private and clinical use.  

\end{abstract}


\newpage
\section{Introduction}

Sleep is an essential biological behavior \cite{baker1985introduction, geyer2009introduction} and therefore highly conserved across animal evolution \cite{joiner2016unraveling}. In mammals, healthy sleep involves a programmed series of characteristic changes in the activity of body and brain, which include alterations of brain wave and breathing patterns, variations of blood pressure, heart beat and body temperature, as well as modulated biochemical activity. Due to the quasi-periodic structure of these changes, they can be viewed as repetitions of a sleep cycle \cite{krauss2021analysis}, subdivided into apparently distinct stages such as Wake, REM, N1, N2 and N3.  

\vspace{0.2cm}\noindent While the recognition of these sleep stages from electroencephalographic (EEG) recordings \cite{barlow1993electroencephalogram} and other measured biosignals was reserved for trained specialists in the past,
the invention of modern machine learning and data analysis tools has quickly led to systems for automatic sleep stage detection \cite{boostani2017comparative, faust2019review, chriskos2021review, krauss2021analysis}, with the promise of eventually freeing medical doctors in sleeping labs from the burden of manual sleep stage classification.

\vspace{0.2cm}\noindent However, the typical automatic classifiers used in the field of machine learning are black boxes with huge numbers of parameters, which makes their classification decisions hard to interpret and difficult to reproduce \cite{castelvecchi2016can, marcus2018deep}. Can these opaque, self-organized, hierarchical features of multi-layer machine learning models be replaced by more transparent, human-interpretable features? 

\vspace{0.2cm}\noindent In a previous paper \cite{metzner2021sleep}, we have demonstrated how to use well-defined statistical operators, such as the mean, the variance, or the kurtosis, for aggregating the raw time series of EEG signals into a few human-interpretable, time-dependent statistical variables. As it turned out, the probability distributions of these variables depend to a certain degree on the sleep stage, and this dependence could be exploited for Bayesian sleep stage detection. In this 'flat' Bayesian approach, all likelihoods and prior probabilities have a simple mathematical meaning, in contrast to the learned weights and biases of a conventional deep multi-layer classifier.  On top of the practical application, our analysis of the time-dependent statistical properties of EEG signals also revealed that certain statistical variables follow continuous trends within and even across the distinct sleep stages - a finding that will be reconfirmed in the present work.

\vspace{0.2cm}\noindent Although automatic sleep stage detection has already been proven possible using a variety of different approaches \cite{boostani2017comparative, faust2019review, chriskos2021review, krauss2021analysis}, the accuracies achieved by these systems are not as satisfactory than in other machine learning applications. Together with the fact that the agreement about sleep stages is relatively poor even among human specialists \cite{fiorillo2019automated}, this raises the question whether the low accuracy merely reflects flaws of the automatic classification systems, or if they are rooted in the data itself: Can human-defined sleep stages be considered as well-defined, 'natural kinds'? Could they be rediscovered by non-supervised data clustering methods without prior knowledge? Are sleep stages really distinct classes or strongly overlapping, fuzzy concepts? What would be the practical consequences of this fuzziness?

\vspace{0.2cm}\noindent We have addressed some of these more theoretical questions in another recent publication \cite{metzner2022classification}. We there argued that classes are in general subjective (user-based) and goal-oriented constructs that must not necessarily reflect the objective structure of the underlying data distributions. Human-defined classes can be useful for certain purposes, even though they do not correspond to well-separated clusters in data space but overlap significantly. This overlap leads however to an upper limit for the achievable classification accuracy \cite{metzner2022classification}.

\vspace{0.2cm}\noindent To test whether the sleep-stages are natural kinds and therefore can be rediscovered by purely objective data analysis, we quantitatively determined the degree of sleep stage clustering in EEG data space, using the previously developed General Discrimination Value (GDV, \cite{schilling2021quantifying}). Finding only an extremely small degree of clustering, even after converting the EEG time series of each epoch to the frequency domain, we next investigated whether the degree of sleep stage clustering can be increased by non-supervised dimensionality reduction with an autoencoder. In this experiment we indeed observed a progressive cluster enhancement over the increasingly narrow autoencoder layers, but quantitatively the effect remained  extremely small \cite{metzner2022classification}.

\vspace{0.2cm}\noindent This work is a follow-up on our former publications regarding the analysis of human EEG signals during sleep \cite{krauss2018analysis,traxdorf2019microstructure,krauss2021analysis, metzner2021sleep,metzner2022classification}, aiming to enhance the degree of sleep stage clustering by suitable ways of pre-processing and dimensionality reduction. As in \cite{metzner2022classification}, we focus on a single EEG channel (the electrode at position F4), split the time series into 30-second epochs that are sleep-stage labeled by a human specialist, and then compute the epoch-wise frequency spectra by Fast Fourier Transformation. Improving upon \cite{metzner2022classification}, we now additional explore the effect of 'scaling' the Fourier amplitudes by taking their modulus to a power of $\gamma$. Each of the resulting epoch-specific 'spectral vectors' is considered as a point in a high-dimensional data space, and we are interested in any kind of cluster structure within this point distribution.

\vspace{0.2cm}\noindent Learning from our past experiments, we now study in advance how the degree of clustering in a data distribution (again quantified by the GDV) depends on the dimensionality of the data space and on the relative fraction of 'separating' and 'non-separating' features/dimensions. This understanding will then inform the optimal selection of data dimensions.  

\vspace{0.2cm}\noindent Since the autoencoder architecture used in \cite{metzner2022classification} for dimensionality reduction of the spectral vectors produced relatively large reconstruction errors, we now turn to Principal Component Analysis (PCA) as an alternative method of 'unsupervised data compression'. This classical method has actually several advantages over neural-network based autoencoders, such as the absence of any tunable model parameters, full mathematical transparency and interpretability, as well as the generation of output dimensions that are mutually uncorrelated.

\vspace{0.2cm}\noindent We then carefully analyze which subset of PCA components should be best included for the GDV-based cluster analysis, comparing different scaling exponents $\gamma$. It turns out that the best cluster separation (with a GDV of -0.211 rather than -0.047 for the uncompressed spectral vectors) is obtained by retaining only the single PCA component $C_1$ with $\gamma=1/2$. According to the 'Eigen-spectrum' of $C_1$, this component basically measures the relative wave contents of the EEG signal in the frequency regimes below 3 Hz (roughly corresponding to the delta-range) and above 3 Hz (covering the theta, alpha, beta and gamma range).

\vspace{0.2cm}\noindent Interestingly, we find an unexpectedly large Pearson correlation coefficient of 0.59 between the time- (or epoch)-dependent PCA component $C_1(t)$ and the 'hypnogram', a common plot of the numerical sleep (or vigilance) label $L(t)$ versus time (with Wake=0, REM=-1, N1=-2, N2=-3 and N3=-4). We verify this result by plotting $C_1(t)$ together with $L(t)$ and conclude that $C_1(t)$ can be regarded as a continuous variable for 'sleep depth', as it indeed closely resembles the hypnogram. Strikingly, this sleep depth variable $C_1(t)$ rises sharply whenever the subject is switching to a more 'shallow' sleep stage, but it falls much more gradually whenever the subject switches to a 'deeper' sleep stage. As  suggested already in \cite{metzner2021sleep}, these results support the idea that sleep is better treated as a continuous process rather than a sequence of distinct stages. 

\vspace{0.2cm}\noindent A practical application of the sleep depth variable $C_1(t)$ is, of course, sleep stage prediction - the automatic sleep stage classification from single-channel EEG data. In contrast to other systems, ours is extremely simple and mathematically fully transparent from the pre-processing up to the final dimensionality reduction method. 

\vspace{0.2cm}\noindent We asses the performance of the method by computing the correlation between the sleep depth variable $C_1(t)$ and the hypnogram, separately for 68 independent full-night sleep recordings, obtaining Pearson correlation coefficients of up to 0.8.

\vspace{0.2cm}\noindent Finally, we test the robustness of the method by recording the same human subject simultaneously with two very different EEG devices and then extract the sleep depth variable $C_1(t)$ separately from both recordings. We find that the two instances of the variable match so closely that they can even be used for a post-hoc temporal synchronization of the EEG machines.


\newpage
\section{Results}

\subsection*{Properties of the Generalized Discrimination Value (GDV)}

In this work, we use the GDV \cite{schilling2021quantifying} as a quantitative measure for cluster separation. Specifically, we investigate to which degree the clustering of data vectors from different sleep stages can be improved by optimizing the parameters of data pre-processing and subsequent unsupervised dimensionality reduction. It is therefore important to know how the GDV depends on the dimensionality of the data space and on the degree of clustering in the individual dimensions of this space.

\begin{figure}[!ht]
\centering
\includegraphics[width=15cm]{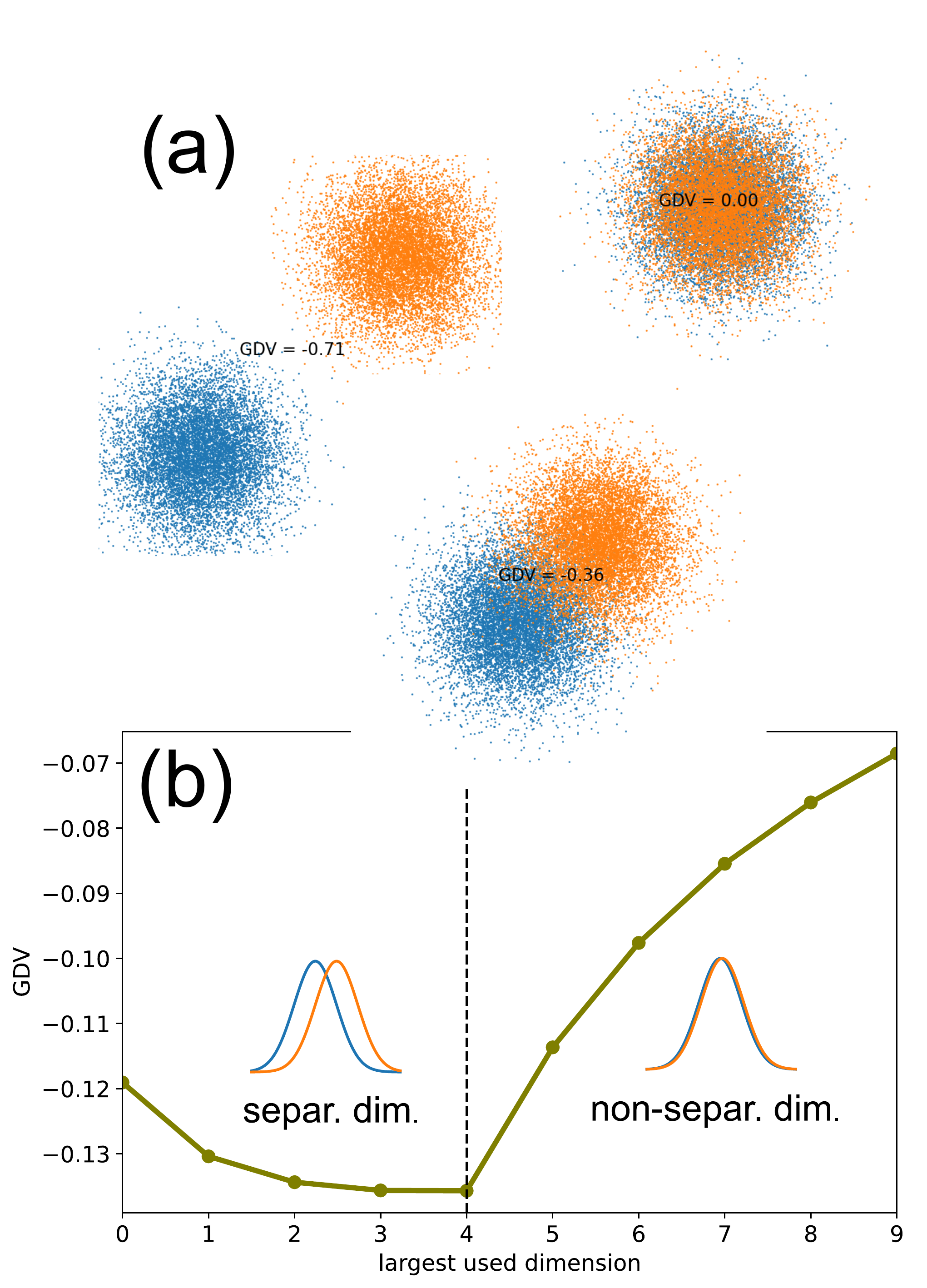}
\caption{Properties of the General Discrimination Value (GDV), illustrated with only two Gaussian clusters (blue and orange). {\bf(a)} Three {\bf examples of cluster pairs} in two dimensions with zero (0.00), medium (-0.36) and large (-0.71) separation. {\bf(b)} {\bf Effect of adding more dimensions} to the GDV cluster evaluation. The first five dimensions 0-4 are (weakly) separating between the two classes and therefore lead to a monotonous decrease of GDV (better clustering), until it saturates. The second five dimensions 5-9 do not separate at all and therefore lead to a monotonous increase of the GDV (worse clustering). It will eventually approach zero (no clustering) if more and more non-separating dimensions are included. This examples demonstrates that non-separating dimensions should be removed from the data. 
\label{fig_GDV}} 
\end{figure}

\vspace{0.2cm}\noindent For this purpose, we generate an artificial ten-dimensional data set with two data classes. The ten components $x_{i=0\ldots 9}$ of the data vectors $\vec{x}$ are mutually independent (as they actually are after a PCA transformation) and normally distributed with unit variance. However, between the two classes the mean values of the Gaussians differ by certain amounts $d_i$. In particular, we assume that $d_i\!=\!1$ (significant separation) for the first five, but $d_i\!=\!0$ (no separation) for the remaining five components. We then compute the GDV of the data set when more and more of the ten components (dimensions) are included (Fig.\ref{fig_GDV}(b)).

\vspace{0.2cm}\noindent We find that the GDV is monotonically decreasing for the first five (class-separating) dimensions, indicating enhanced clustering of the data points, but the GDV already shows a clear saturation in this example. Consequently, if several dimensions are available in which the data classes are well separating, it can be beneficial to include more than one of these separating dimensions.

\vspace{0.2cm}\noindent However, once the non-separating dimensions are subsequently included, the GDV is strongly increasing, indicating a progressive loss of the clustering that was already achieved before. As a general rule, it is therefore important to remove all non-separating dimensions from the data space, before the clustering structure of the data points is analyzed.

\subsection*{Enhancing sleep-stage clustering}

\begin{figure}[!ht]
\centering
\includegraphics[width=16cm]{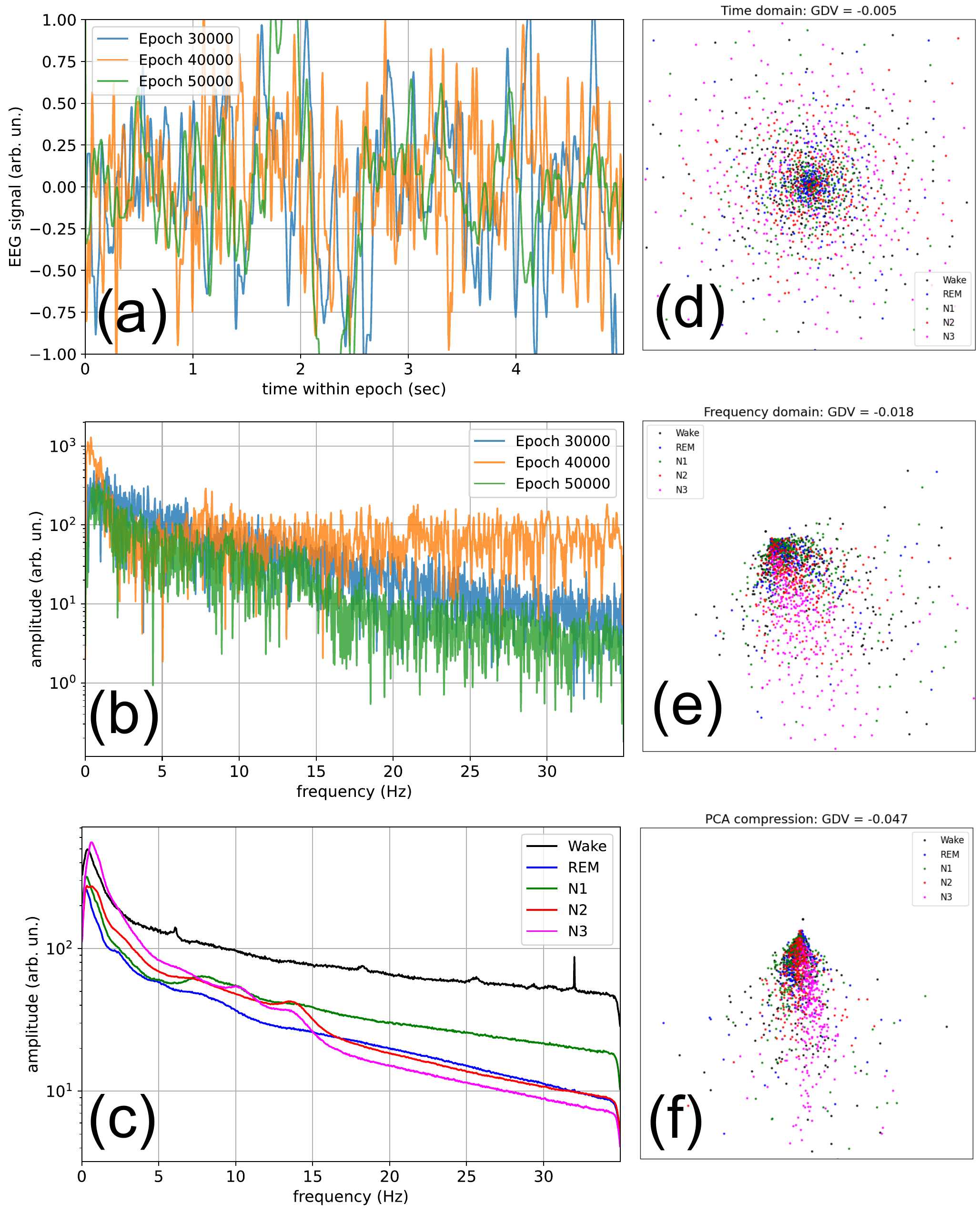}
\caption{Raw data, pre-processing, and clustering. 
{\bf(a)} {\bf Time-domain EEG signals} during the first five seconds of three arbitrary recorded epochs.
{\bf(b)} {\bf Frequency-domain EEG spectra} (Modulus of the complex FFT amplitudes) for the same three epochs.
{\bf(c)} {\bf Average EEG spectra} in each of the five sleep stages.
{\bf(d)} Two-dimensional MDS visualization of the epoch-wise {\bf signal vectors}. For the plot we randomly selected 1500 examples of these 7680-dimensional time-domain vectors (300 from each sleep stage). The GDV value of -0.005 indicates virtually no clustering of the five sleep stages.
{\bf(e)} MDS visualization of the epoch-wise {\bf spectral vectors}. For the plot we randomly selected 1500 examples of these 1050-dimensional frequency-domain vectors.  The GDV of -0.018 indicates a very weak clustering.
{\bf(f)} MDS visualization of the epoch-wise {\bf compressed vectors}, using only the 5 lowest PCA components. For the plot we randomly selected 1500 examples of these 5-dimensional vectors. In this case, the GDV of -0.047 indicates a clustering that is significantly better than in the cases (d) and (e).
\label{fig_FFT}} 
\end{figure}

Next, we apply the GDV to quantify the degree of sleep stage clustering in 70174 thirty-second long epochs of recorded one-channel EEG signals (For details about the data and pre-processing see Methods section. Three example epochs are shown in Fig.\ref{fig_FFT}(a)). 

\vspace{0.2cm}\noindent Each of these 'signal vectors' can be considered as a point in a 7680-dimensional space and has been labeled by a sleep specialist to belong to one of the five sleep stages (Wake, REM, N1, N2, and N3). The distribution of these points in their data space can be visualized in only two dimensions by using Multi-Dimensional Scaling (MDS, cf. Methods). The MDS visualization of the time-domain signal vectors shows no hints of clustering (Fig.\ref{fig_FFT}(d)), and this is confirmed by a deneralized discrimination value (GDV, cf. Methods) of only -0.005.

\vspace{0.2cm}\noindent Next we transform the data vectors to Fourier space, keeping only the first 1050 of the real-valued frequency-dependent amplitudes (Three example epochs are shown in Fig.\ref{fig_FFT}(b)). The MDS visualization of the resulting spectral vectors shows now already a small degree of clustering (Fig.\ref{fig_FFT}(e)), corresponding to a GDV value of -0.018.

\vspace{0.2cm}\noindent The small clustering in Fourier space means that the spectral vectors are significantly different in the five sleep stages. This is confirmed by computing the average frequency spectra for each sleep stage (Fig.\ref{fig_FFT}(c)). Note that the high-frequency components are progressively suppressed in the 'deeper' sleep stages. Only the spectrum of the REM phase (blue) does not follow this general ordering.

\vspace{0.2cm}\noindent Next, we perform a PCA-based dimensionality reduction with the spectral vectors, keeping only the first five PCA components (For the MDS visualization, see Fig.\ref{fig_FFT}(f)). This results in a significant enhancement of the sleep stage clustering, yielding a GDV value of -0.047.

\subsection*{Optimal subset of PCA components}

\begin{figure}[!ht]
\centering
\includegraphics[width=16cm]{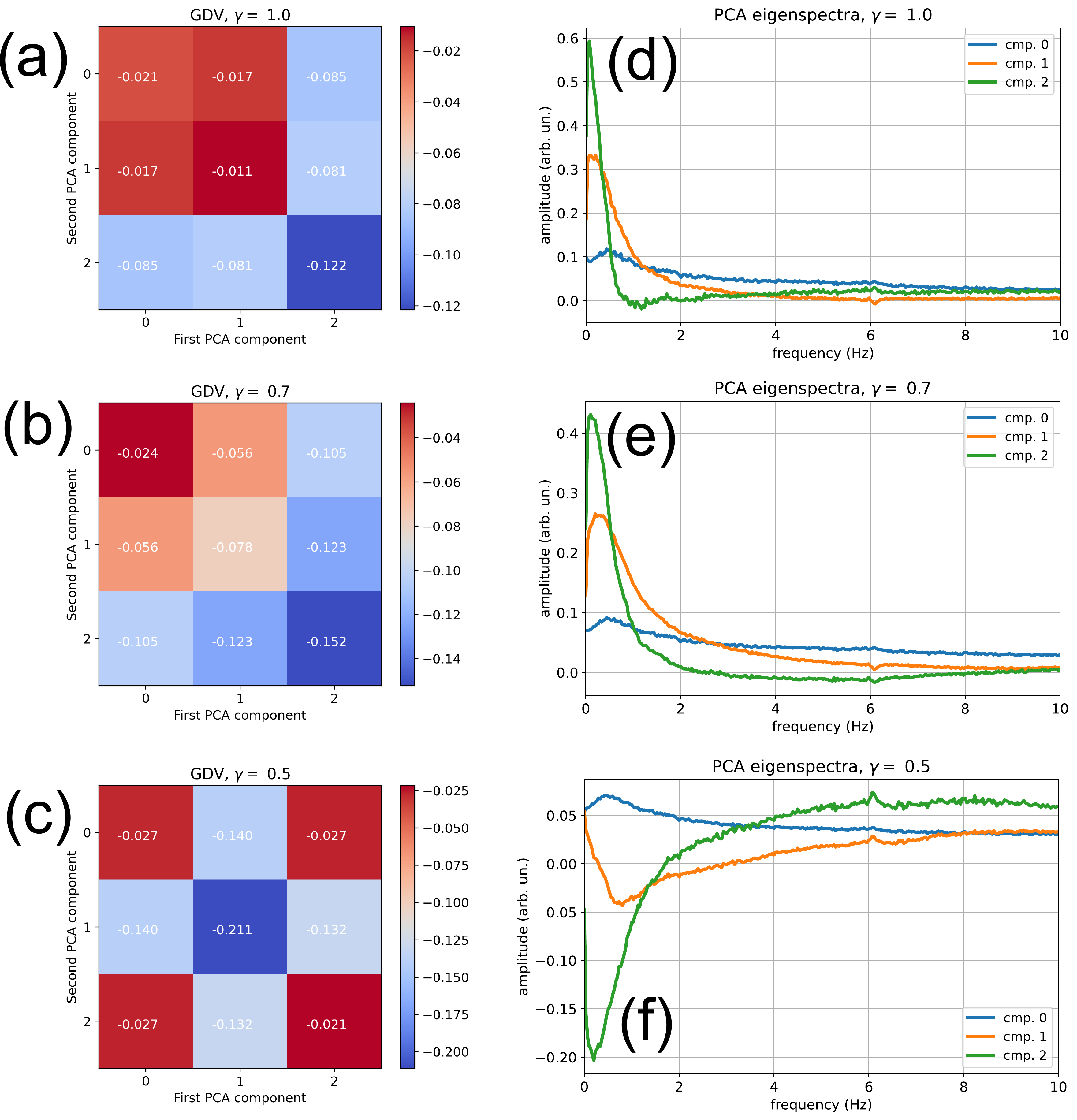}
\caption{Separability of sleep stages and characteristic spectra of the first three PCA components $k\in\left\{0,1,2\right\}$, evaluated for different re-scaling exponents $\gamma$. The matrices (a-c) show the General Discrimination Values (GDV) when different pairs of PCA components are used. The values on the matrix diagonal correspond to the use of a single component only. The characteristic spectra (d-f) for each PCA component $k$ show which frequencies are enhanced (positive values) or reduced (negative values) relative to the average spectrum of the data set, if component $k$ is set to one and all other components to zero.
{\bf(a,d)} For $\gamma=1.0$, sleep stages separate best (GDV = -0.122) if only PCA-component $k=2$ is used. This component is large if small frequencies below about 1 Hz are dominant in the Fourier spectra (positive part of the green curve in d).
{\bf(b,e)} For $\gamma=0.7$, the situation is similar to the case of $\gamma=1.0$, but now the maximally separating component $k=2$ corresponds to a large spectral content in the frequency range from 0 to about 2 Hz.
{\bf(c,f)} For $\gamma=0.5$, sleep stages separate best (GDV = -0.211) if only PCA-component $k=1$ is used. The larger this component, the more high frequencies above $\approx$ 3 Hz dominate over lower frequencies (orange curve in f) in the EEG signal.
\label{fig_matspec}} 
\end{figure}

We have shown above that in order to enhance the degree of clustering in a multi-dimensional data space, only the class-separating dimensions should be retained. For this purpose, we next investigate how the GDV is changing when only one or two out of the first three PCA components are selected. The results can be presented in the form of a symmetric matrix, where the diagonal elements correspond to only one selected component (See heat maps in Fig.\ref{fig_matspec}(a-c)). In addition, we investigate the effect of changing the scaling exponent $\gamma$ of the spectral vectors.

\vspace{0.2cm}\noindent For $\gamma=1$, the best cluster separation is achieved when only the PCA component $C_2$ is kept (Fig.\ref{fig_matspec}(a)). The GDV in this case is -0.122. For $\gamma=0.7$, the best cluster separation is again achieved with only the PCA component $C_2$ (Fig.\ref{fig_matspec}(b)). The GDV is now -0.152. For $\gamma=0.5$, however, the best cluster separation is achieved with only $C_1$ (Fig.\ref{fig_matspec}(c)). The GDV is then -0.211, indicating a quite significant degree of clustering. Note that values of $\gamma$ smaller than 1/2 can lead to detrimental effects in the subsequent analysis.

\begin{figure}[!ht]
\centering
\includegraphics[width=18cm]{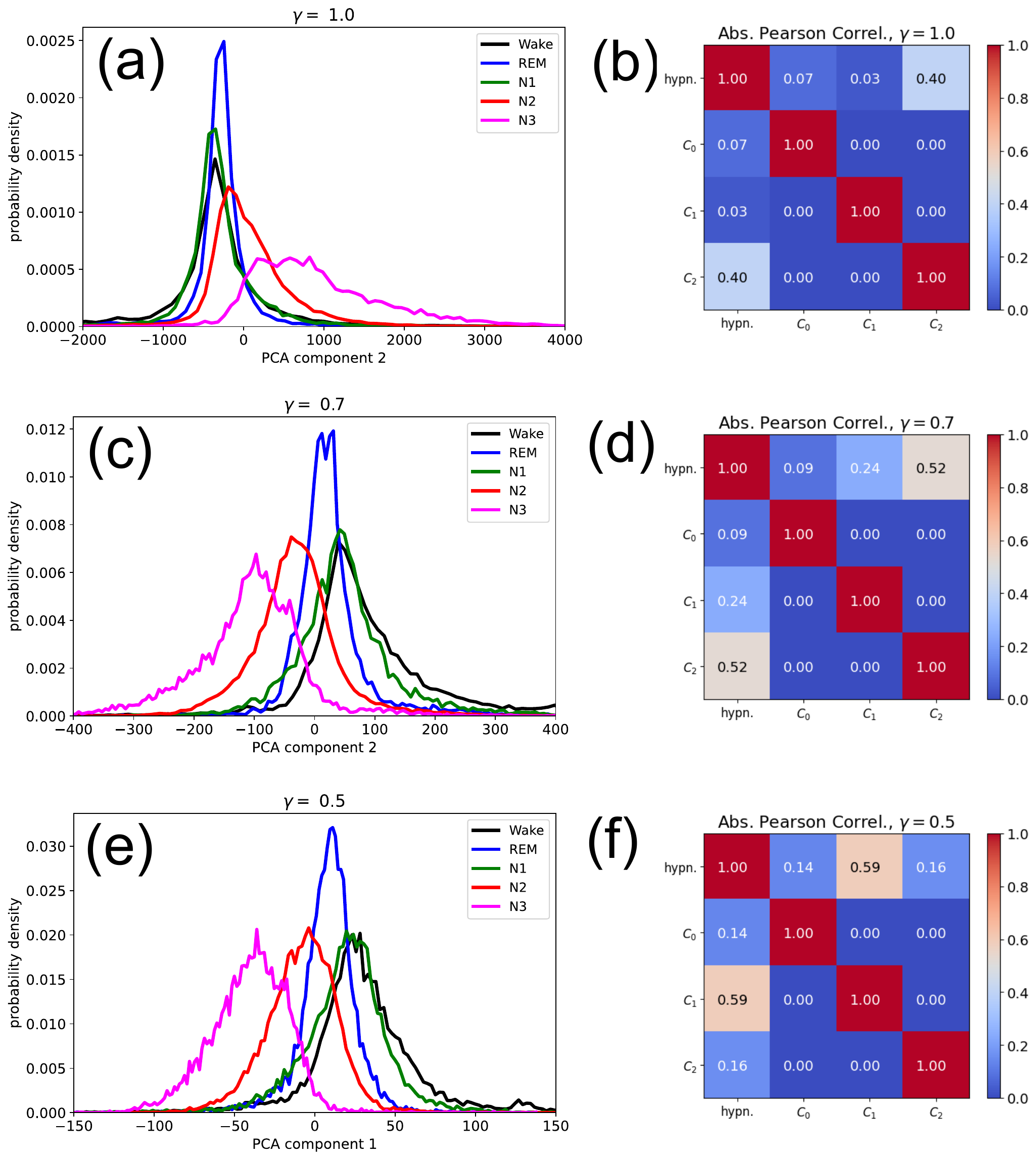}
\caption{{\bf Left column}: Probability distributions of the best separating PCA component in the five sleep stages, {\bf Right column}: (Absolute value of the) Pearson correlations between the hypnogram and the first three PCA components. Results are shown for different rescaling exponents $\gamma$ (rows). {\bf(a,b)} For $\gamma=1.0$, component $C_2$ is maximally separating, but the stages Wake, REM and N1 can be hardly distinguished. The correlation between $C_2$ and the hypnogram is 0.4. {\bf(c,d)} For $\gamma=0.7$, component $C_2$ is again separating best. Now all sleep stages, with the exception of N1 (green) and Wake (black), can be distinguished. The correlation between $C_2$ and the hypnogram is 0.52. {\bf(c)} For $\gamma=0.5$, the superior component is $C_1$. Graphically, the separability of the sleep stages appears similar to (b), but actually the GDV is significantly smaller (compare Fig.\ref{fig_matspec}). The correlation between $C_1$ and the hypnogram is 0.59. Note that, in all cases, the PCA-component with the best cluster separation is also correlating best with the hypnogram. 
\label{fig_Distr}} 
\end{figure}

\vspace{0.2cm}\noindent The quantitative GDV values above can also be visually confirmed by plotting the distributions of the single best-separating PCA components in the five sleep stages (Fig.\ref{fig_Distr}(a,c,e)). Note that the maxima of the peaks in the distributions are at different positions for the five sleep stages (with the exception that Wake (black) and N1 (green) are surprisingly similar), but nevertheless there is a significant overlap between all five peaks. This is the kind of 'fuzziness' mentioned in the Introduction, which eventually limits the achievable accuracy in automatic sleep stage classifiers.

\subsection*{Eigenspectra of the PCA components}

The spectral vectors form a complex distribution of points in their data space. If this distribution, for the sake of clarity, is imagined as a spheroidal point cloud in a three-dimensional space, the center of mass of this point cloud corresponds to the mean EEG frequency spectrum $V_{mean}(f)$, averaged over all recorded epochs. In this image, the PCA is finding the main axes of the spheroid (the orthogonal axes of maximum variation in the data distribution), each corresponding to one principal component $i$. It places into data space a new coordinate system that consists of these main axes, with the origin located at the center of the point cloud. Moving from the origin into the direction of one of the main axes $i$ by an amount $C_i$ means to modify the frequency spectrum away from the average in a well-defined way: The modification can be mathematically expressed by adding a perturbation to the mean spectrum, namely $C_i$ times the 'eigenspectrum' $\Delta V_i(f)$ of PCA component $i$. In general, a point in data space with PCA coordinates $\vec{C}=(C_1,C_2,\ldots)$ corresponds to the frequency spectrum
\begin{equation}
V(\vec{C},f) = V_{mean}(f) + \sum_{i\!=\!0}^{N_c}\; C_i\cdot\Delta V_i(f).
\end{equation}

\vspace{0.2cm}\noindent In Fig.\ref{fig_matspec}(d,e,f), we plot the Eigenspectra for the first three PCA components, and for the three tested values of the scaling exponent $\gamma$. In the case of $\gamma=0.5$, the Eigenspectrum for the best separating component $C_1$ (orange curve in Fig.\ref{fig_matspec}(f)) is negative for small frequencies between zero and about 3 Hz, and positive for larger frequencies. Consequently, the more negative PCA component $C_1$ becomes, the more the low-frequency brain waves dominate over the high frequency waves in the EEG spectrum. We therefore expect $C_1$ to become more negative in the 'deeper' sleep stages.

\subsection*{Correlation of PCA components and sleep labels}

In order to test whether some PCA components indeed correlate with sleep depth, we first compute the mutual Pearson correlation coefficients between the components $C_0\ldots C_2$ and the negative numerical sleep label (Wake=0, REM=-1, N1=-2, N2=-3 and N3=-4), which corresponds to a  hypnogram when plotted over time (more precisely: over successive sleep epochs $k$). Since positive and negative correlations are equally interesting here, we only consider the modulus of the Pearson coefficients. They are again presented in the form of a symmetrical matrix in Fig.\ref{fig_Distr}(b,d,f). 

\vspace{0.2cm}\noindent The mutual correlations between $C_0$, $C_1$ and $C_2$ are zero, as it should be in PCA. More interestingly, the correlation between the hypnogram and the three PCA components (first row of each matrix) is non-zero, and it is maximal for the same PCA component that also gives the best cluster separation. In particular, for the a scaling exponent of $\gamma=0.5$, the Pearson coefficient between $C_1$ and the hypnogram reaches a surprisingly large value of 0.59. We therefore consider only the case $\gamma=0.5$ for the rest of this work.

\subsection*{Sleep depth variable $C_1(t)$ versus hypnograms}

\begin{figure}[!ht]
\centering
\includegraphics[width=13cm]{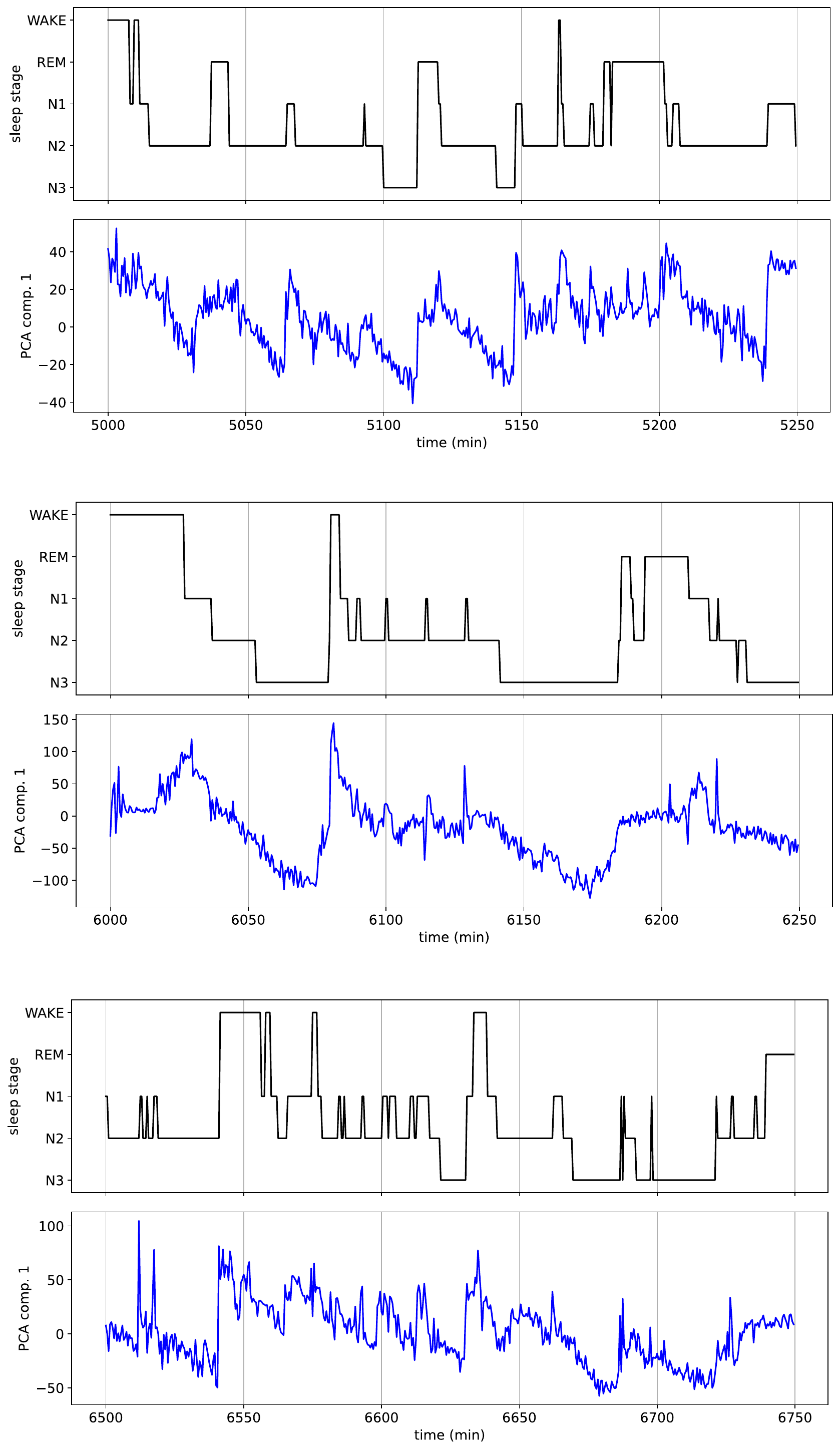}
\caption{{\bf Temporal evolution of the PCA component $C_1(t)$} for $\gamma=0.5$ (blue), {\bf and its relation to the human-generated hypogram} (black). Shown are three time intervals with a duration of 250 minutes each. Note that $C_1(t)$ tracks the general 'depth of sleep' very well. However, where the hypnogram is jumping abruptly to one of the deeper sleep stages, $C_1(t)$ shows a gradual decrease. 
\label{fig_Hypno}} 
\end{figure}

To further test the relation between $C_1(t)$ and the hypnogram, we plot both quantities during the same time period. Fig.\ref{fig_Hypno} shows three arbitrarily chosen time periods, each with a duration of 250 minutes. It is obvious that $C_1(t)$ (blue) resembles the hypnogram (black) quite closely, but there are also some characteristic differences: Whenever the sleep label switches upwards, to a more shallow sleep stage or to the Wake state, $C_1(t)$ also shows an abrupt increase. By contrast, when the label switches downwards, to a deeper sleep stage, $C_1(t)$ decreases as well, but in a very gradual way that can easily continue for more than 30 minutes. These results confirm that (for $\gamma=1/2$), the PCA component $C_1(t)$ indeed encodes sleep depth in a continuous way. 

\subsection*{Performance of $C_1(t)$ in individual data sets}

For potential future applications of $C_1(t)$ in personal sleep tracking devices and other low-cost scenarios, it is important to analyze the degree of correlation between the sleep depth variable and the ground truth hypnogram for a larger group of full-night recordings. Moreover, it is unclear whether a better performance of $C_1(t)$ can be achieved when the PCA is fit 'globally' to all available EEG data sets, or 'locally' to each individual subject.

\begin{figure}[!ht]
\centering
\includegraphics[width=16cm]{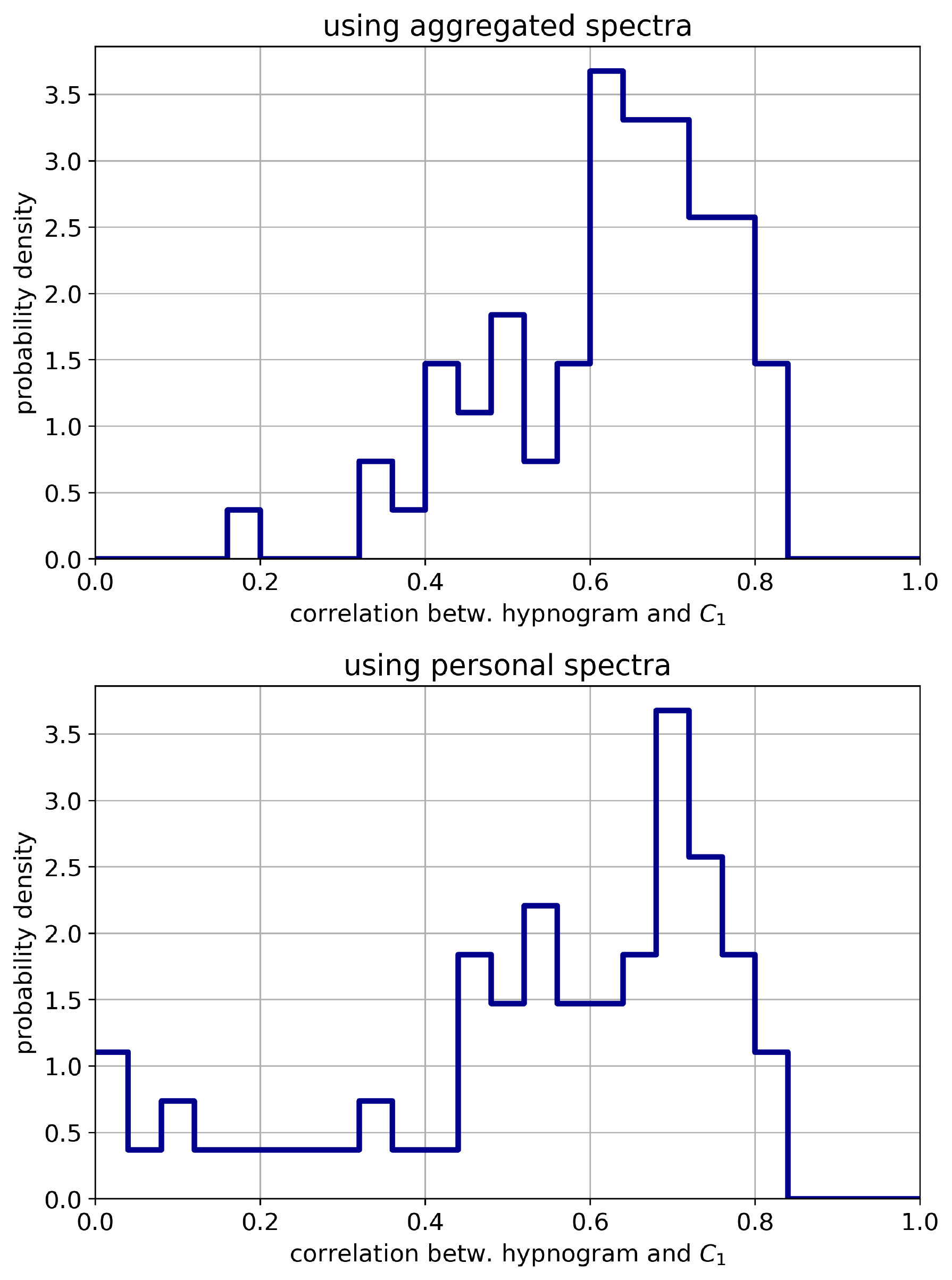}
\caption{ 
Distribution of (absolute) Pearson correlation coefficients between the hypnogram and PCA component $C_1$. {\bf Top}: Distribution when the PCA is fit to the aggregated spectra from all 68 data sets. {\bf Bottom}: Distribution when the PCA is fit to the personal spectrum only. 
\label{fig_Personal}} 
\end{figure}

\vspace{0.2cm}\noindent To answer these questions, we compute the Pearson correlation coefficient between $C_1(t)$ and the hypnogram separately for all 68 available EEG data sets. The results are shown as probability distributions in Fig.\ref{fig_Personal}.

\vspace{0.2cm}\noindent We find a surprisingly large fraction of Person correlations in the 'good' range
$\left[ 0.6,0.8 \right]$ and in the 'medium' range $\left[ 0.4,0.6 \right]$, no matter if the PCA is fitted globally (top panel) or locally (bottom panel). However, using only person-specific information for the PCA fit results in a significantly larger fraction of 'bad' outcomes in the range $\left[ 0,0.4 \right]$. It is therefore more appropriate to fit the PCA model to a large pool of full-night EEG recordings before the sleep depth variable is evaluated for a new individual.

\subsection*{Sleep depth from different EEG devices}

Since the degree of sleep stage clustering was shown in this work to depend strongly on the used data pre-processing (including parameters such as the spectral scaling exponent $\gamma$), it is likely that using different EEG devices with distinct spectral responses and noise levels will also affect the clustering and eventually might lead to non-comparable time courses of the extracted sleep depth variable $C_1(t)$.

\vspace{0.2cm}\noindent To address this potential problem, we measure the same human subject simultaneously with two different EEG setups. One of them is a clinical device with only three EEG channels and the other one is a 64-channel 'research' device, but we are using only the channel 'F4' for our comparison. The two machines are not mutually synchronized or coupled in any way, and they are run at different sampling rates.

\begin{figure}[!ht]
\centering
\includegraphics[width=16cm]{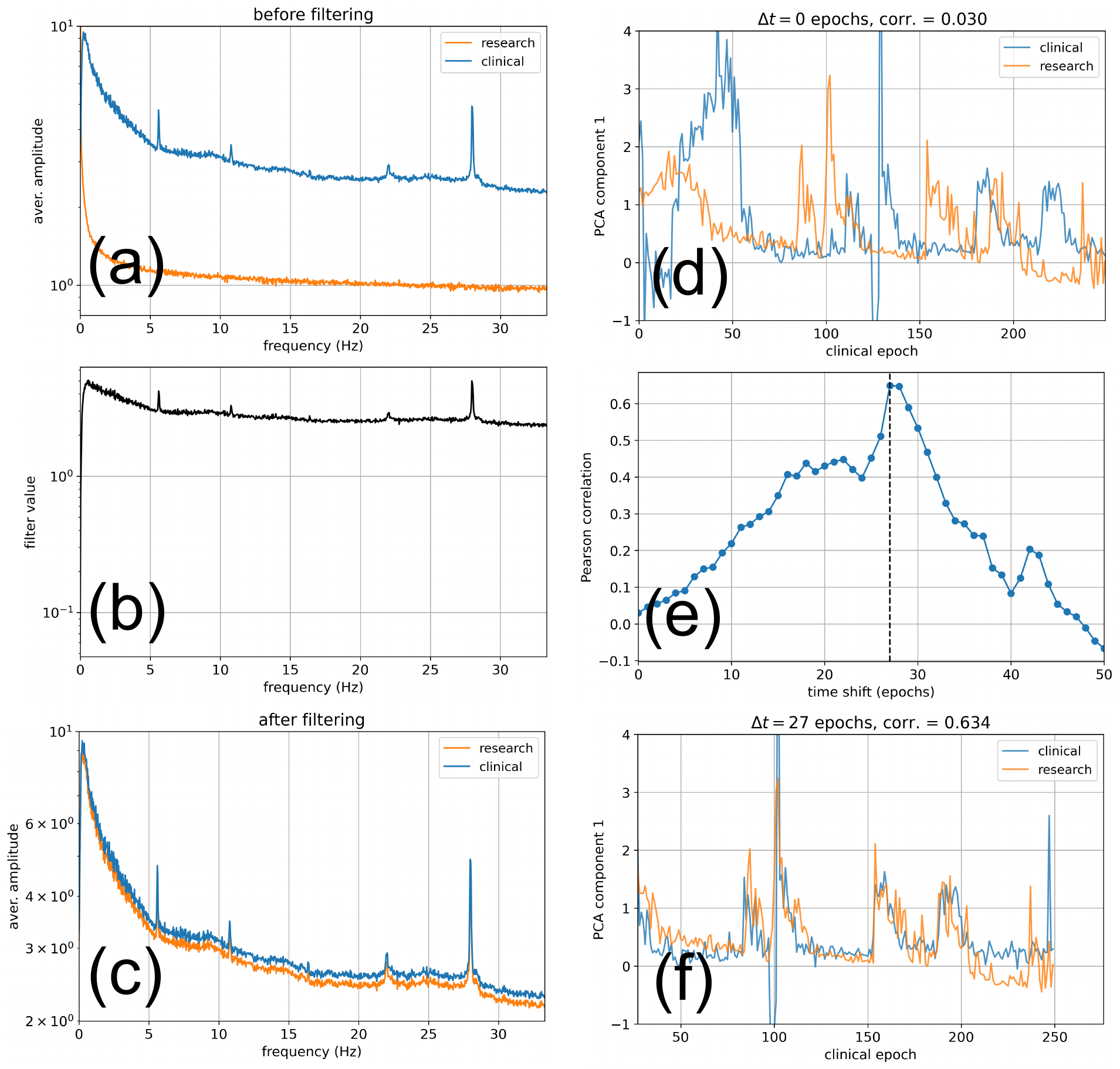}
\caption{ 
{\bf Comparing PCA component $C_1$ derived from different EEG devices}. The same sleeping subject is simultaneously measured with a 3-channel 'clinial' device (blue) and and 64-channel 'research' device (orange). We focus on channel F4 in both cases. By averaging the spectral vectors over all 30-second epochs, we find that the two devices produce different frequency spectra {\bf (a)}. To compensate for these differences, we multiply the spectra of the research device with a suitable filter function {\bf (b)}. After filtering, the two devices produce the same average spectra {\bf (c)}. Since the two EEG devices are not synchronized and have been activated at different times, the time dependence of component $C_1(t)$ during the first 250 epochs appears different {\bf (d)}. The Pearson correlation between $C_1(t)$ from the two devices however changes when a time shift is artificially introduced {\bf (e)}. The correlation has a clear maximum for a shift of $\Delta t=27$ epochs. When this optimal time shift is applied, the time courses $C_1(t)$ from the two devices clearly match {\bf (f)}. The PCA component $C_1$ is therefore a robust measure that can even be used and compared even across devices. 
\label{fig_Align}} 
\end{figure}

\vspace{0.2cm}\noindent After recording a whole night of sleep, the F4-signal of the research device is down-sampled to the sampling frequency of the clinical device (128 Hz). When computing the Fourier spectra (re-scaled with $\gamma=1/2$), averaged over all sleep epochs, we find that the spectral responses of the two devices are quite different (Fig.\ref{fig_Align}(a)). To make the data as similar as possible, we compute a filter function in frequency space, defined as the ratio between the average spectral responses of the clinical and research device (Fig.\ref{fig_Align}(b)). By multiplying each epoch-specific spectral vector of the research device with this filter function, the two responses become identical on average (Fig.\ref{fig_Align}(c)).

\vspace{0.2cm}\noindent Next, we fit a PCA model to the spectral vectors of the clinical machine. Note that this pool of spectral vectors comes only from a single full-night recording. In future practical applications, it might be preferable to use a larger pool of recordings instead.

\vspace{0.2cm}\noindent Based on the given PCA model, we now extract the sleep depth variable $C_1(t)$ from both machines (Fig.\ref{fig_Align}(d)). As expected, the time courses do not match, because the two devices are not synchronized and have actually been switched to recording mode at different times.

\vspace{0.2cm}\noindent However, by artificially applying different time-shifts $\Delta t$ (multiples of 30-second epochs) between the two $C_1(t)$ signals and computing the Pearson correlation coefficient for each $\Delta t$, we find that this cross-correlation has a clear global peak at 27 epochs (Fig.\ref{fig_Align}(e)).

\vspace{0.2cm}\noindent Shifting the $C_1(t)$ time coarse of the research device by this optimum amount, we find a remarkably close match between the two extracted sleep variables (Fig.\ref{fig_Align}(f)). It is therefore possible to directly compare the $C_1(t)$ signals measured with different EEG devices by using the above technique. 

\vspace{0.2cm}\noindent Another interesting future application would be to compute $C_1(t)$ separately for the different electrodes of a given EEG device and then to investigate if different brain regions 'fall asleep' at slightly different times or to different degrees.

\newpage
\section{Discussion}

In this work, we continue a line of investigation that focuses on the cluster structure of human EEG data during different sleep stages. We start with the assumption that if sleep stages would be 'natural kinds', they should be discoverable by completely unsupervised methods of data analysis as distinct clusters in data space. However, using the General Discrimination Value as a quantitative measure of class separability, we only find a vanishingly small degree of clustering (GDV=-0.005) in the raw epoch-wise EEG signals. This is not surprising, because two signal vectors that have been shifted relative to each other by only a few time steps can have an arbitrarily large euclidean distance, even if they belong to the same sleep stage. In general, finding any class-separating features in time-domain EEG data is extremely difficult for a machine learning system without prior information. Even if some method of unsupervised data analysis could find well-separated clusters directly in the space of time-domain data vectors, it is doubtful whether those will correspond to the human-defined sleep stages. 

\vspace{0.2cm}\noindent For this reason, we believe that a minimal human-assisted pre-processing and feature selection is required for the analysis of EEG sleep data, an approach that might be called 'weakly supervised' data analysis. In the present study, Fourier-transforming the epoch-wise time-domain vectors to the frequency domain and subsequently ignoring all phase information was sufficient to improve sleep-stage clustering significantly (GDV=-0.018), and a further improvement was possible by scaling the resulting frequency spectra by a suitable exponent $\gamma$. It is however likely that sleep-stage clustering could be further enhanced in future studies by choosing different time windows (currently we have restricted our numerical experiments to 30-second epochs only), or by replacing the Fourier transformation by a suitable wavelet transformation. 

\vspace{0.2cm}\noindent We could confirm in the present study that a substantial enhancement of clustering (up to GDV=-0.211) is possible by a suitable dimensionality reduction of the pre-processed data, provided that only the best separating dimensions (features) are retained. Surprisingly, the classical method of Principal Component Analysis (PCA) proved to be much better suited for compressing Fourier-EGG data than a formerly tested multi-layer auto-encoder.   
\vspace{0.2cm}\noindent In particular, we found that the single PCA component $C_1(t)$, in combination with a spectral scaling exponent of $\gamma=1/2$, not only maximizes sleep-stage clustering, but also correlates surprisingly well (Pearson correlation of 0.59) with the hypnogram, a traditional plot of the numerical sleep stage label over time (epoch). We note that this correlation might be further enhanced by adjusting the arbitrary numerical values assigned to the five different sleep stages.

\vspace{0.2cm}\noindent An analysis of the 'eigenspectrum' of $C_1(t)$ reveals that this PCA component basically measures the relative content of slow (below 3 Hz) and fast (above 3 Hz) brain waves in the momentary spectrum of the EEG signal. It is remarkable that this well-known discriminating feature emerged as being optimal for enhancing sleep-stage clustering in our weakly supervised approach.

\vspace{0.2cm}\noindent When plotting the time-course of $C_1(t)$ together with the hypnogram, we find that the two curves are strikingly similar, so that $C_1(t)$ can actually be viewed as a sleep depth variable. However, the two quantities differ in their behavior at transitions from one discrete sleep stage to another: When the brain is 'switching' (according to the criteria of the human rater) to a more shallow sleep stage or to the wake stage, $C_1(t)$ is also abruptly rising. By contrast,  switches to deeper sleep stages only mark the beginning of a continuous downward trend of $C_1(t)$ that can last for more than 30 minutes. During these downward trends, the EEG's frequency spectrum is gradually becoming dominated by slow brain waves,  even though the sleep stage is rated as constant. 

\vspace{0.2cm}\noindent In any case, the gradual evolution of the brain wave spectrum within a fixed sleep stage, together with the strongly overlapping probability distributions of $C_1$ in the various sleep stages, strongly suggest that sleep is better understood as a continuum, rather than a succession of discrete phases. The sleep depth variable $C_1(t)$ offers an easy, transparent and reproducible way to track this continuous brain process based on single channel EEG data.


\newpage
\section{Methods}

\subsection*{Three-channel sleep EEG data}

For the main part of this paper, we are using 68 three-channel EEG data sets from the sleep laboratory of University Hospital Erlangen, each corresponding to a full-night recording of brain signals from a different human subject. The data were recorded with a sampling rate of 256 Hz, using three separate channels F4-M1, C4-M1, O2-M1. In this work, however, we are only using the first channel (F4-M1). 

\vspace{0.2cm}\noindent  The participants of the study included 46 males and 22 females, with an age range between 21 and 80 years. Exclusion criteria were a positive history of misuse of sedatives, alcohol or addictive drugs, as well as untreated sleep disorders. The study was conducted in the Department of Otorhinolaryngology, Head Neck Surgery, of the Friedrich-Alexander University Erlangen-Nürnberg (FAU), following approval by the local Ethics Committee (323–16 Bc). Written informed consent was obtained from the participants before the cardiorespiratory polysomnography (PSG). 

\vspace{0.2cm}\noindent  After recording, the raw EEG data were analyzed by a sleep specialist accredited by the German Sleep Society (DGSM), who detected typical artifacts \cite{tatum2011artifact} in the data and visually identified the five sleep stages (Wake, REM, N1, N2, N3) in subsequent 30-second epochs, according to the AASM criteria (Version 2.1, 2014) \cite{iber2007aasm,american2012aasm}.

\subsection*{Multi-dimensional scaling (MDS)}

A frequently used method to generate low-dimensional embeddings of high-dimensional data is t-distributed stochastic neighbor embedding (t-SNE) \cite{van2008visualizing}. However, in t-SNE the resulting low-dimensional projections can be highly dependent on the detailed parameter settings \cite{wattenberg2016use}, sensitive to noise, and may not preserve, but rather often scramble the global structure in data \cite{vallejos2019exploring, moon2019visualizing}.
In contrast to that, multi-Dimensional-Scaling (MDS) \cite{torgerson1952multidimensional, kruskal1964nonmetric,kruskal1978multidimensional,cox2008multidimensional} is an efficient embedding technique to visualize high-dimensional point clouds by projecting them onto a 2-dimensional plane. Furthermore, MDS has the decisive advantage that it is parameter-free and all mutual distances of the points are preserved, thereby conserving both the global and local structure of the underlying data. 

When interpreting patterns as points in high-dimensional space and dissimilarities between patterns as distances between corresponding points, MDS is an elegant method to visualize high-dimensional data. By color-coding each projected data point of a data set according to its label, the representation of the data can be visualized as a set of point clusters. For instance, MDS has already been applied to visualize for instance word class distributions of different linguistic corpora \cite{schilling2021analysis}, hidden layer representations (embeddings) of artificial neural networks \cite{schilling2021quantifying,krauss2021analysis}, structure and dynamics of recurrent neural networks \cite{krauss2019analysis, krauss2019recurrence, krauss2019weight}, or brain activity patterns assessed during e.g. pure tone or speech perception \cite{krauss2018statistical,schilling2021analysis}, or even during sleep \cite{krauss2018analysis,traxdorf2019microstructure}. 
In all these cases the apparent compactness and mutual overlap of the point clusters permits a qualitative assessment of how well the different classes separate.

\subsection*{Generalized Discrimination Value (GDV)}

We used the GDV to calculate cluster separability as published and explained in detail in \cite{schilling2021quantifying}. Briefly, we consider $N$ points $\mathbf{x_{n=1..N}}=(x_{n,1},\cdots,x_{n,D})$, distributed within $D$-dimensional space. A label $l_n$ assigns each point to one of $L$ distinct classes $C_{l=1..L}$. In order to become invariant against scaling and translation, each dimension is separately z-scored and, for later convenience, multiplied with $\frac{1}{2}$:
\begin{align}
s_{n,d}=\frac{1}{2}\cdot\frac{x_{n,d}-\mu_d}{\sigma_d}.
\end{align}
Here, $\mu_d=\frac{1}{N}\sum_{n=1}^{N}x_{n,d}\;$ denotes the mean, and $\sigma_d=\sqrt{\frac{1}{N}\sum_{n=1}^{N}(x_{n,d}-\mu_d)^2}$ the standard deviation of dimension $d$.
Based on the re-scaled data points $\mathbf{s_n}=(s_{n,1},\cdots,s_{n,D})$, we calculate the {\em mean intra-class distances} for each class $C_l$ 
\begin{align}
\bar{d}(C_l)=\frac{2}{N_l (N_l\!-\!1)}\sum_{i=1}^{N_l-1}\sum_{j=i+1}^{N_l}{d(\textbf{s}_{i}^{(l)},\textbf{s}_{j}^{(l)})},
\end{align}
and the {\em mean inter-class distances} for each pair of classes $C_l$ and $C_m$
\begin{align}
\bar{d}(C_l,C_m)=\frac{1}{N_l  N_m}\sum_{i=1}^{N_l}\sum_{j=1}^{N_m}{d(\textbf{s}_{i}^{(l)},\textbf{s}_{j}^{(m)})}.
\end{align}
Here, $N_k$ is the number of points in class $k$, and $\textbf{s}_{i}^{(k)}$ is the $i^{th}$ point of class $k$.
The quantity $d(\textbf{a},\textbf{b})$ is the euclidean distance between $\textbf{a}$ and $\textbf{b}$. Finally, the Generalized Discrimination Value (GDV) is calculated from the mean intra-class and inter-class distances  as follows:
\begin{align}
\mbox{GDV}=\frac{1}{\sqrt{D}}\left[\frac{1}{L}\sum_{l=1}^L{\bar{d}(C_l)}\;-\;\frac{2}{L(L\!-\!1)}\sum_{l=1}^{L-1}\sum_{m=l+1}^{L}\bar{d}(C_l,C_m)\right]
 \label{GDVEq}
\end{align}

\noindent whereas the factor $\frac{1}{\sqrt{D}}$ is introduced for dimensionality invariance of the GDV with $D$ as the number of dimensions.

\vspace{0.2cm}\noindent Note that the GDV is invariant with respect to a global scaling or shifting of the data (due to the z-scoring), and also invariant with respect to a permutation of the components in the $N$-dimensional data vectors (because the euclidean distance measure has this symmetry). The GDV is zero for completely overlapping, non-separated clusters, and it becomes more negative as the separation increases. A GDV of -1 signifies already a very strong separation.

\subsection*{Generic pre-processing of EEG epochs}

Each recorded single-channel epoch consists of $30\times256=7680$ EEG signal values, and our database of 68 person-specific data sets contains totally more than 70000 of these labeled epochs. In the following, we denote the signal value $U$ at time step $t\in\left\{1,\ldots,7680\right\}$ within epoch $e$ of the personal data set $s\in\left\{1,\ldots,68\right\}$ by $U^{(s)}_{e,t}$.

\vspace{0.2cm}\noindent Within each of the 68 data sets, we first remove all epochs in which artifacts where detected by the sleep specialist. The remaining epochs are normalized by z-scoring, 
\begin{equation}
U^{(s)}_{e,t} \longrightarrow 
\frac{U^{(s)}_{e,t} - \mu^{(s)}}{\sigma^{(s)}},
\end{equation}
where $\mu^{(s)}$ is the average signal value and $\sigma^{(s)}$ is the standard deviation in data set $s$. By this way, variations of the signal amplitudes between different subjects $s$ are suppressed, but variations between the sleep stages of each individual subject are retained.

\vspace{0.2cm}\noindent Pooling now over all subjects $s$, we create a unified list of normalized {\bf signal vectors} $U_k(t)$. Each of them corresponds to one of the global epochs $k\in\left\{1,\ldots,70174\right\}$, contains time steps $t\in\left\{1,\ldots,7680\right\}$, and has a known sleep-stage label $L_{k}$.

\vspace{0.2cm}\noindent Next, we convert the time-domain signal vectors $U_k(t)$ to the frequency domain by Fast Fourier Transformation (FFT), yielding 3840 complex Fourier amplitudes $\hat{A}_k(f)$ for each epoch $k$. Discarding the phase information, we retain only the real-valued modulus, which is taken to the power of $\gamma$ (called the scaling exponent) in order to enhance cluster separation later on. This produces a total number of 70174 {\bf spectral vectors}, defined as $V_k(f)=\left|\hat{A}_k(f) \right|^{\gamma}$, representing the scaled momentary frequency spectrum during the global epoch $k$. Since our EEG device produces a strong drop of all frequency components above about 35 Hz, we keep only the entries below this cutoff, so that $f\in\left\{1,\ldots,1050\right\}$.

\vspace{0.2cm}\noindent Finally, we perform a Principal Component Analysis (PCA) on the list of spectral vectors $V_k(f)$, from which we keep only the first three components $C_0$, $C_1$ and $C_2$. By projecting all spectral vectors into this three-dimensional subspace, we obtain 70174 {\bf compressed vectors} $W_k(C)$, each containing the 'most essential information' about the momentary frequency spectrum in a given epoch.

\vspace{0.2cm}\noindent Each of the 70174 data vectors (either the signal, spectral, or compressed vectors) can be considered as a point in a vector space. In the following, we are interested in the distribution of these points, and we analyze how well they cluster with respect to the five human-defined sleep stages.

\subsection*{Sleep depth variable $C_1(k)$}

It turns out that the best separation of sleep stages is obtained with the first PCA component $C_1 = \mbox{PCA}_1$ and when using the scaling exponent $\gamma=0.5$ (compare Results section). Since the temporal development of $C_1$ over subsequent sleep epochs $k$ closely resembles the hypnogram, it can be interpreted as a 'sleep depth' variable. Summing up, it is computed from the temporal EEG signal $U_k(t)$ in epoch $k$ in the following way:
\begin{equation}
C_1(k) = \mbox{PCA}_1\left\{ 
\left|\;
\mbox{FFT}\left\{
U_k(t)
\right\}
\;\right|^{1/2}
\right\}.
\end{equation}

\subsection*{Sleep depth $C_1(k)$ from different EEG devices}

For testing the robustness of $C_1(k)$, we use a separate overnight data set from a different human subject, recorded simultaneously with two EEG devices in a sleeping lab of the Paracelsus Medical University, Nürnberg. The first ('clinical') device has 3 channels and was set to a sampling frequency of 128 Hz. The second ('research') device has 64 channels and was set to a larger sampling frequency, but has after the measurement been down-sampled to 128 Hz as well. In the following, we use only the signals of channel $F4$ from both devices. 

\vspace{0.2cm}\noindent As before, the EEG signals are split into subsequent 30-second epochs $k$, yielding the signal vectors $U_k(t)$. The latter are Fourier transformed, keeping only the lowest 1000 frequency components, and then scaled with the exponent $\gamma=1/2$, yielding the spectral vectors

\begin{equation}
V_k(f) = 
\left|\;
\mbox{FFT}\left\{
U_k(t)
\right\}
\;\right|^{1/2}.
\end{equation}

\vspace{0.2cm}\noindent By averaging these spectral vectors over all epochs $k$, we obtain two overall frequency spectra, $\hat{V}^{(cli)}(f)$ and $\hat{V}^{(res)}(f)$, one for each device (See Fig.\ref{fig_Align}(a)). As they are too different for a direct comparison, we compute a frequency-dependent filter function as the ratio between the clinical and research frequency spectra (See Fig.\ref{fig_Align}(b))
\begin{equation}
A(f) = \frac{\hat{V}^{(cli)}(f)}{\hat{V}^{(res)}(f)}. 
\end{equation}
We now multiply each of the epoch-specific spectral vectors of the research device with this filter function
\begin{equation}
V_k^{(res)}(f)\;\longrightarrow\;
V_k^{(res)}(f)\;A(f).
\end{equation}

\vspace{0.2cm}\noindent After this filtering, the overall (epoch-averaged) frequency spectra $\hat{V}^{(cli)}(f)$ and $\hat{V}^{(res)}(f)$ of the two devices become identical (See Fig.\ref{fig_Align}(c), where a small difference has been artificially introduced for better visibility).

\vspace{0.2cm}\noindent Next, we fit a PCA model to the spectral vectors $V^{(cli)}(f)$ of the clinical device. This single model is then used, for both devices, to compress the 1000-dimensional spectral vectors $V_k(f)$ down to two-dimensional vectors $W_k(C) = \left( C_0(k),C_1(k) \right)$. We further consider only the PCA component $C_1(k)$, because it can be interpreted as a variable for sleep depth (See Fig.\ref{fig_Align}(d,e)).


\section*{Additional Information}

\subsection*{Author contributions}

CM conceived the study, implemented the methods, evaluated the data, and wrote the paper. PK co-designed the study, discussed the results and wrote the paper. AS, HS and KT discussed the results. MT provided data.

\subsection*{Funding}
This work was funded by the Deutsche Forschungsgemeinschaft (DFG, German Research Foundation): grant SCHU\,1272/16-1 (project number 455908056) to HS, grant TR\,1793/2-1 (project number 455908056) to MT, grant SCHI\,1482/3-1 (project number 451810794) to AS, and grants KR\,5148/2-1 (project number 436456810), KR\,5148/3-1 (project number 510395418) and GRK\,2839 (project number 468527017) to PK.

\subsection*{Competing interests statement}
The authors declare no competing interests.

\subsection*{Data availability statement}
The complete data and analysis programs will be made available upon reasonable request.

\subsection*{Ethical approval and informed consent}
The main part of the study (68 three-channel data sets) was conducted in the Department of Otorhinolaryngology, Head Neck Surgery, of the Friedrich-Alexander University Erlangen-Nürnberg (FAU), following approval by the local Ethics Committee (323 – 16 Bc). Written informed consent was obtained from the participants before the cardiorespiratory poly-somnography (PSG). The comparison between different EEG machines (64-channel and 3-channel device) was conducted in the Department of Otorhinolaryngology, Head and Neck Surgery, Paracelsus Medical University, Nürnberg, Germany, following approval by the local Ethics Committee (103 – 20 B). Written informed consent was obtained from the participants before the cardiorespiratory poly-somnography (PSG).

\subsection*{Third party rights}
All material used in the paper are the intellectual property of the authors.


\bibliographystyle{unsrt}
\bibliography{references}


\clearpage

\clearpage

\clearpage

\clearpage

\clearpage

\clearpage

\clearpage

\end{document}